\newif\ifsubmode
\submodefalse

\ifsubmode
  \documentclass[12pt,preprint]{aastex}
  \received{}
  \revised{}
  \accepted{}
\else
  \documentclass[12pt,preprint]{emulateapj}

\fi

\newenvironment{inlinefigure}{
\def\@captype{figure}
\noindent\begin{minipage}{0.999\linewidth}\begin{center}}
{\end{center}\end{minipage}\smallskip}

\newcommand{\figonecap}{
 Variance of the cosmic shear (filled squares) from the MDS
catalog for different sizes of the averaging window.  Open circles
indicate the spurious signal that could arise from known PSF errors;
crosses the variance inferred from simulated catalogs in the absence of
shear.  Error bars on the measured points represent 1-$\sigma $ errors
estimated from simulations; the error bars on the other points represent
the error on the mean.
}

\newcommand{\figtwocap}{
 Collation of cosmic shear measurement from sources quoted in the text,
plus Maoli et al.~(2001) and van Waerbeke et al.~(2001).
The results presented here and those of Refregier et al.~(2002) are from
HST data; all others are from ground-based data.  The quoted scale is
that of a square region, with circular regions converted to the square
area of the same size.  Error bars include both systematic and
statistical errors, as applicable.
}

\slugcomment{To appear in  {\it Astrophysical Journal Letters}, December 1, 2003} 
\shorttitle{Cosmic shear from MDS}
\shortauthors{Casertano, Ratnatunga, Griffiths}

\begin{document}

\title{Cosmic Gravitational Shear from the HST Medium Deep Survey}

\author{Stefano Casertano}
\affil{Space Telescope Science Institute, 3700 San Martin Drive, 
Baltimore, MD 21218}

\and

\author{Kavan U. Ratnatunga and Richard E. Griffiths}
\affil{Physics Dept., Carnegie Mellon University, Pittsburgh, PA 15213}

\email {stefano@stsci.edu; kavan, griffith@astro.phys.cmu.edu}

\begin{abstract}

We present a measurement of cosmic shear on scales ranging from $
10\arcsec $ to $ 2\arcmin $ in 347 WFPC2 images of random fields.  Our
result is based on shapes measured via image fitting and on a simple
statistical technique; careful calibration of each step allows us to
quantify our systematic uncertainties and to measure the cosmic shear
down to very small angular scales.  The WFPC2 images provide a robust
measurement of the cosmic shear signal decreasing from $5.2\% $ at $
10\arcsec $ to $2.2\% $ at $ 130\arcsec $.

\end{abstract}


\section {INTRODUCTION}

Gravitational lensing may be in principle the most powerful tool to
measure the distribution of dark matter from galactic to cosmic
scales. Lensing can be used to directly trace the gravitational
potential of all matter, regardless of its nature, and only requires
the distribution of light sources which shine through the matter
distribution---unlike velocity measurements, which depend on the
availability of visible tracers {\it in situ}. Considerable progress
has been made in recent years in the use of weak lensing to trace dark
matter associated with light, on scales ranging from individual
galaxies (Brainerd et al.~1996, Griffiths et al.~1996, Blandford et
al.~2001) to
clusters of galaxies (e.g., Mellier 1999, Kaiser 2001).
On even larger scales, gravitational lensing can trace directly the
distribution and clustering of dark and luminous matter via its
gravitational signature, cosmic shear. In principle, cosmic shear can
provide an unbiased measurement of the spectrum of mass fluctuations,
independent of their nature and of any light associated with them.
Theoretical calculations predict a signature of a few percent on 
arcminute scales for several models of structure formation
and cosmology (Jain and Seljak 1997, Kaiser 1998, Bartelmann
\& Schneider 1999, Barber 2002).

Measurements of the cosmic shear have indeed been achieved over the last
few years.  A tentative detection was announced by our group in July
1999 (published in Casertano et al.~2001).  Shortly afterwards, several
ground-based programs have achieved more reliable detections of cosmic
shear at a level of 1\% to 3\% on angular scales ranging from $ 1\arcmin
$ to $ 15\arcmin $ (van Waerbeke et al.~2000, Kaiser et al.~2000, Bacon
et al.~2000, Wittman et al.~2000).  These measurements placed
significant, if preliminary, constraints on the strength of mass
clustering at $ z \sim 0.5 $ on linear scales from 0.2 to 5 Mpc (Kaiser
et al.~2000).  In the last year, the second generation of ground-based
weak shear measurement has started to probe the power spectrum of dark
matter fluctuations and constrain cosmological parameters (Bacon et
al.~2003, Brown et al.~2003, van Waerbecke et al.~2002).  Somewhat weaker
constraints have been obtained from space-based data by Refregier et
al.~(2002) and Rhodes et al.(2001), who use data similar to ours.

Typically, ground-based results are based on measuring the shape of
marginally resolved galaxies over very large areas---up to several
square degrees. The quality of the results depends critically on the
correction for a number of systematic noise sources, which for
ground-based data include atmospheric effects as well as those due to
variations in the Point Spread Function (PSF) across the field of view
and any other image distortion, due to either optics or sensitivity
variations. Very sophisticated methods have been developed to extract
this signal optimally (e.g., Kaiser et al.~1995, Kaiser 2000, Bernstein and Jarvis 2002).
Nonetheless, ground-based measurements could be affected to some
extent by residual systematics, and independent confirmation from
different methods on other types of data, such as those obtained with
HST, is valuable. With HST, individual galaxies are substantially
larger than both PSF and pixel size, and therefore instrumental
effects, while present, are both smaller than and different from
those found in ground-based data.

The present paper offers a measurement of cosmic shear using 347
HST WFPC2 images collected by the Medium-Deep Survey (MDS), covering a
total area of 0.53 square degrees. The measurement is based on a new
statistical technique, described and tested in detail in a companion
paper (Ratnatunga et al.~2003, hereafter RCG), and takes full
advantage of the high resolution and relative stability of the WFPC2
PSF. Our method takes properly into account observational and PSF
errors, as well as statistical effects that become important on small
angular scales; unique to our procedure is the ability to measure the
cosmic shear signal from the distribution of observed galaxy shapes
even on scales that include only a few galaxies per cell. In
consequence, we measure cosmic shear over angular scales ranging from
the full field of view of the camera ($ \sim 2 \arcmin $) down to $
10$--$20\arcsec $, using approximately 35,000 galaxies. The
procedures we develop can be applied to other space-based samples with
a larger number of galaxies, and will probably allow an improvement of
the quality and reliability of cosmic shear measurements with the
upcoming trove of Advanced Camera for Surveys data.

\section{The Data: the MDS catalog of galaxies}
\label{sec:data}

The HST MDS has resulted in a large catalog of
galaxies suitable for weak lensing measurements.  Extensive tests and
simulations (Ratnatunga et al.~1999, hereafter RGO) show that the
parameters---size, shape, magnitude---measured for such galaxies are
unbiased and have quantifiable, reliable error estimates.  The MDS
catalog covers several hundred fields observed with the Wide Field and
Planetary Camera 2 (WFPC2); this study includes 347 fields covering a
total area of 0.53 square degrees to a typical depth of $ I = 25.5 $.
The object catalog for each field is produced using the standard MDS
procedures, i.e., automated object location, eyeball verification,
two-dimensional image fitting, and goodness-of-fit classification.  We
use the galaxy parameters as determined by RGO, with a small
correction for the image shear introduced by the geometric distortion
of the WFPC2.  The full catalog for the 347 selected fields contains
36,389 stars and 100,253 galaxies, as classified by our likelihood
ratio algorithm.

We consider only objects classified as galaxies with a signal-to-noise
parameter $ \Xi > 1.6 $, as defined in RGO.  We further exclude about
500 very small objects with half-light radius $ r_h < \min (0\farcs05,
1\arcsec \times 10^{-2.4+0.5\Xi} $.  Applying these selections leaves a total of
52,433 confirmed galaxies with $ 21 < I < 26 $, of which about 35,000
are used in the actual shear measurements. 

\section {Measuring Cosmic Shear: Galaxy Parameters and the Excess
Quadratic Ellipticity Statistic} 
 \label {sec:statistic}

Current measurements of cosmic shear are based on the correlated
distortion of background galaxy images.  The measurement consists of two
steps: first, determine the {\it corrected} shape of the galaxies
detected in those images, after accounting for instrumental and seeing
effects; second, combine these single-galaxy measurements into a
statistical measure of the cosmic shear. 

Our shape measurements are based on the axis ratio and position angle
determined in the MDS pipeline by fitting the observed image with a
parametric light distribution, convolved with the PSF.  RGO describe
the fitting process in detail and provide a reliable estimate of the
uncertainties in the derived image parameters for each galaxy. For
ideal data, the axis ratio and position angle thus defined transform
correctly under the effect of shear, and therefore are suitable for
weak lensing measurements.  Unlike most ground-based observations, the
galaxies are generally well resolved in WFPC2 images, and the effect
of the WFPC2 PSF is a small correction---about a few percent---on the
measured image parameters.  In Section~\ref{sec:problems} we report on
our tests of the impact of non-ideal data.

For the second step, we have developed a statistical method based on the
{\it excess quadratic ellipticity} $ A_f = \langle e_i \rangle^2 -
\langle e_i^2 \rangle / N_g $, which measures the correlation between
the ellipticities $ e_i $ of the $ N_g $ galaxies within each cell.  If
the galaxies are randomly oriented, the $ e_i $ are uncorrelated and $
A_f $ will vanish on average.  A non-random component, in the form of a
correlation between observed galaxy ellipticities, will cause a slight
positive bias of $ A_f $, which scales quadratically with the shear. 
Typically this bias is too small to be measured in a single cell;
averaging a large number of cells measures the variance of the
cell-averaged shear.  The bias in $ A_f $ is especially difficult to
measure when $ N_g $ is small; partly for this reason, most studies
report only shear on relatively large scales, $ 1\arcmin $ or more. 
However, as we show in RCG, careful analysis of the statistical
properties of the excess quadratic ellipticity enables us to measure the
typical shear down to scales of $ 10\arcsec $, where $ N_g \lesssim 10
$.  In RCG we show that optimal results are obtained by using
appropriate weights for each field, and we find that for our data, the
relationship between $ A_f $ and the shear variance is approximately:
 \begin{equation}\label{eq:Af calibration}
    \langle A_f \rangle \sim (3.82 - 2.10/N_g - 3.13 \langle e^2 \rangle
       - 7.35 \langle \sigma_e^2 \rangle ) \langle \gamma^2 \rangle
 \end{equation}
 where $ \langle \gamma^2 \rangle $ is the variance in the cosmic
shear averaged in each cell, $ \langle e^2 \rangle $ is the mean
squared ellipticity of all galaxies, and $ \sigma_e $ is the combined
measurement error in each ellipticity component, estimated by
combining the measurement error in both axis ratio and position angle.
Since $ \sigma_e $ is typically much smaller than the ellipticity
itself, this latter term can usually be neglected.  Note that the
relationship between $ A_f $ and $ \langle \gamma^2 \rangle $ depends
significantly on the number of galaxies for $ N_g \ll 10 $, and this
dependence must be taken into account when averaging results for small
cell sizes.

Our method is in principle quite similar to other methods that rely on
estimating the correlations between ellipticities of galaxies that are
proximate on the sky; see, e.g., Seitz et al.~(1998) and Rhodes et
al.~(2000).  Our key improvements are 1) considering explicitly the
number of galaxies per cell, as in Equation~\ref{eq:Af calibration}
above, which allows us to obtain unbiased shear estimates for small
cells including only a handful of galaxies each, and 2) using
extensive simulations to validate the statistical properties of our
method, which allow us to derive the uncertainty in our measurements
with high fidelity.  More details are reported in RCG.

\section {Cosmic shear from the MDS data}
\label {sec:measurement}

 \subsection{Measured shear}

Figure~\ref{fig:final shear} shows the cosmic shear estimated on
scales from $ 10\arcsec $ to $ 130\arcsec $ on the WFPC2 data.
Cells larger than $ 65\arcsec $ are L-shaped as the
WFPC2 field of view, with a gap of 10--30$ \arcsec $ between the
detectors to avoid edge effects.  For homogeneity, we do not include
PC data in our analysis.

The measured variance of the cosmic shear, represented by the filled
squares, is based on the calibration of the $ A_f $ statistic given in
Equation~\ref{eq:Af calibration}.  The error bars reflect the
1-$\sigma$ uncertainty in the shear variance determined by the
simulations described in RCG; we find that the presence
of shear increases only slightly the measurement uncertainty, and
therefore the 1-$\sigma$ errors in Figure~\ref{fig:final shear}
apply to a null measurement as well.  The open circles show
the estimated contribution of PSF errors, while the crosses indicate
the inferred shear variance from simulations in which no shear was
included.  Error bars in this case show the uncertainty in the mean
effect, based on several thousand simulations, rather than the
dispersion in individual simulations.

Our data yield the strongest evidence of cosmic shear on a
single-detector scale, $ 65\arcsec $, where the measured value $ \langle
\gamma^2 \rangle = 0.0007 $ is about 4-$\sigma $ from a null result. 
The measured shear increases on smaller scales, although individual
measurements have lower significance; the detection at $ 10\arcsec $ has
a significance of about 2-$\sigma$.  

 \ifsubmode
  \placefigure{fig:final shear}
 \else
 \begin{inlinefigure}
 \begin{center}
 \resizebox{\textwidth}{!}{\includegraphics{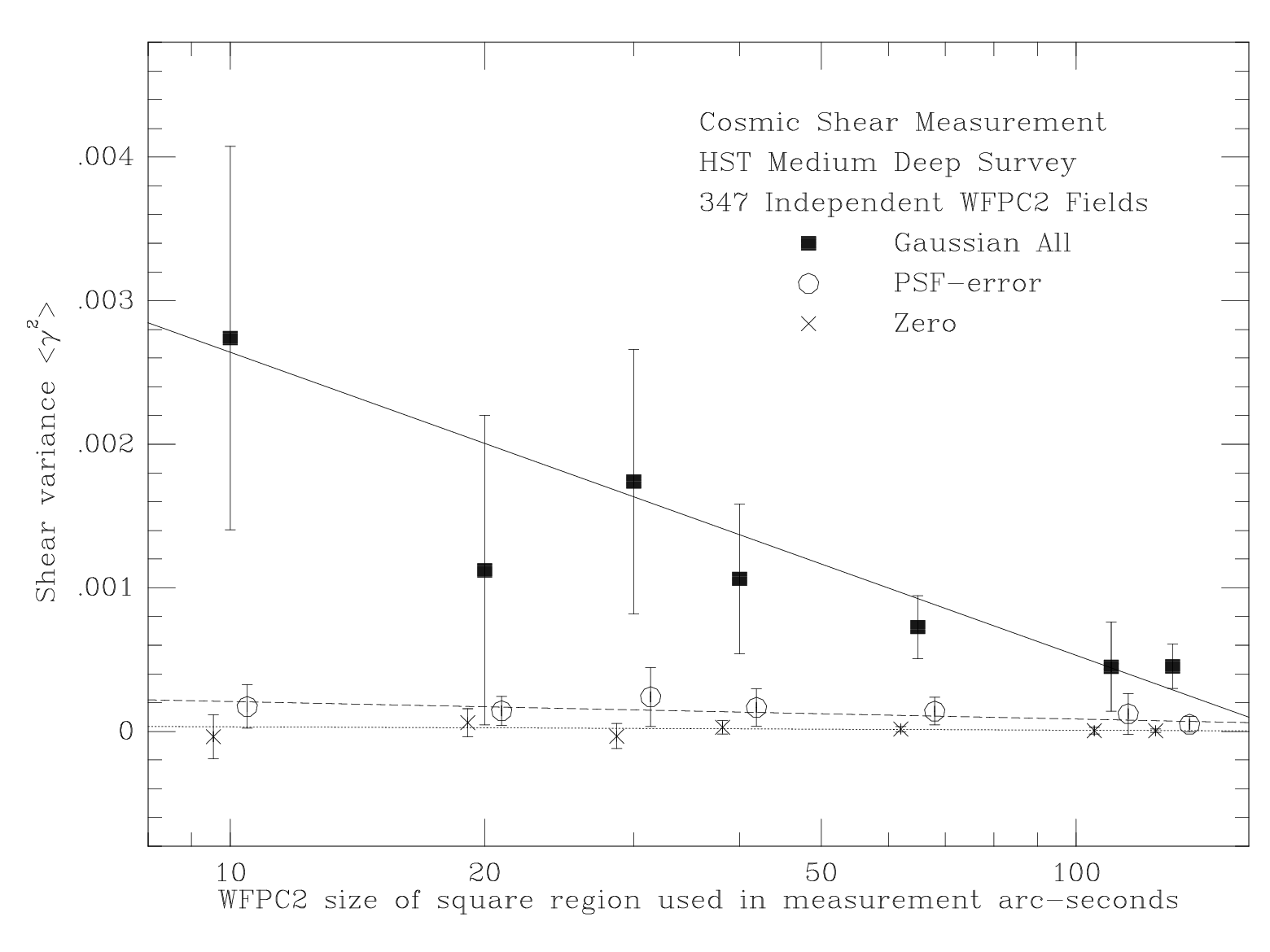}}
 \end{center}
 \figcaption{\figonecap
 \label{fig:final shear}}
 \end{inlinefigure}
 \fi

 \subsection{Possible sources of false signal}
\label{sec:problems}

The image fitting method we adopt to determine galaxy shapes and
orientations has been shown (RGO) to yield measurements of galaxy
properties that are not affected by detectable systematic biases, even
in the presence of image pixelation.  However, the cosmic shear
measurement is sensitive to other effects due to an imperfect
knowledge of the observations.  We discuss briefly three
such effects: errors in the adopted PSF, uncorrected differential
aberration in parallel HST observations, and fields with special
properties, such as unknown galaxy clusters.

WFPC2 has a difficult-to-characterize PSF with significant variations
over its field of view. The PSF is temporally stable except for its
dependence on the HST focus, which is known to vary systematically
over a single orbit as well as over longer time scales. However, the
focus dependence affects primarily the core of the PSF; our tests show
that even a substantial focus error affects the shape of a typical
galaxy in our sample, as measured via the MDS procedures, by much
less than 0.1\%.  Of more concern is a systematic mismatch between the
true PSF and the model we adopted, which is based on the TinyTim
code (Krist 2000).  If improperly accounted for, the intrinsic
asymmetry of the PSF could produce an artificial correlation between
measured ellipticities and thus affect our cosmic shear measurement.
We test for such a correlation in two ways.  First, we generate artificial
catalogs in which each galaxy retains its observed size, orientation,
and position in detector coordinates, but is assigned randomly to
another observation. When analyzed exactly as the true catalog, 
the artificial catalogs yield a very small residual shear, less than 
0.03\% on all scales, as shown by the open circles
in Figure~\ref{fig:final shear}.  This test is
also sensitive to other persistent instrumental errors, such as
the assumed geometric distortion.  Second, we compute the mean
ellipticity of the galaxies that fall in each 100$\times$100 pixel
region of the WFPC2 field of view in all observations.  A significant
non-zero mean would indicate residual instrumental effects, such
as uncorrected PSF; we find instead a typical rms value of 
2.9\% for each ellipticity component, compared to an expected
value of $ 3.0 \pm 0.2\% $ from the same number of randomly oriented
objects in each cell.  It is worth noting that an earlier version of
the MDS catalog, based on a 3x3 grid of PSF models for each detector
instead of the 7x7 grid used here, fails this test, returning a mean
ellipticity $ \sim 1\% $ with an rms variation of $ 3.8\% $.

Differential velocity aberration can affect parallel observations
because the HST pointing is corrected for velocity aberration at the
location of the primary target of the observation.  As a consequence,
parallel observations are convolved with a potentially asymmetric
kernel which is generally less than $ 0\farcs01 $, but can occasionally
be as large as $ 0\farcs 035 $ (Ratnatunga et al.~1995).  The
effect of velocity aberration is expected to have two characteristic
signatures: as a convolution, it should affect small galaxies more
than large ones; and its impact should vary depending on the location
of the primary target of the observation.  We find no significant
variation in our results as a function of either galaxy size or
location of the primary target, suggesting that differential velocity
aberration is not significant within our measurement accuracy.
More accurate measurements of cosmic shear, based on many more galaxies
than we have considered here, might require a correction for this effect.

Large concentrations of mass, such as galaxy clusters, produce a shear
signature much larger than its typical value.  Truly random
observations should include galaxy clusters as often as they occur
over the whole sky; {\it on average,} the cosmic shear variance
measured via random observations should be an unbiased representation
of the variance of cosmic shear.  However, a small number of
cluster-affected fields can skew the measured shear significantly.  We
have verified that the distribution of $ A_f $ measured from
individual pointings is regular and is not dominated by a small number
of outlying points.  To be safe, we have excluded two fields with
somewhat larger than normal values of $ A_f $, without any measurable
effect on the derived value of the shear variance.  We conclude that
our results are not dominated by a small number of fields with high
correlation.

\section {Discussion}

The cosmic shear we measure for the $ \sim 0.5 $ square degrees
covered by the MDS data is generally consistent with the measurements
obtained from other data, as well as with the results of the Refregier
et al.\ (2002) analysis of essentially the same observations.
Our procedures are substantially different from those of other
authors: we measure image parameters via image fitting, including PSF
convolution, and follow a different, carefully calibrated analysis
process (see RGO and RCG for more details).  Thanks to our methods, we
can measure cosmic shear on very small scales, down to $ 10\arcsec $,
into the regime of a very small number of galaxies per cell.  On scales
where our measurements overlap those of other authors, we find a
somewhat larger cosmic shear signature; for example, we find a value
of $ \gamma $ about 30\% larger than Refregier et al.\ (2002) measured
on very similar data, and the difference is about twice the combined
1--$\sigma$ uncertainty.

 \ifsubmode
  \placefigure{fig:various shear measurements}
 \else
 \begin{inlinefigure}
 \begin{center}
 \resizebox{\textwidth}{!}{\includegraphics{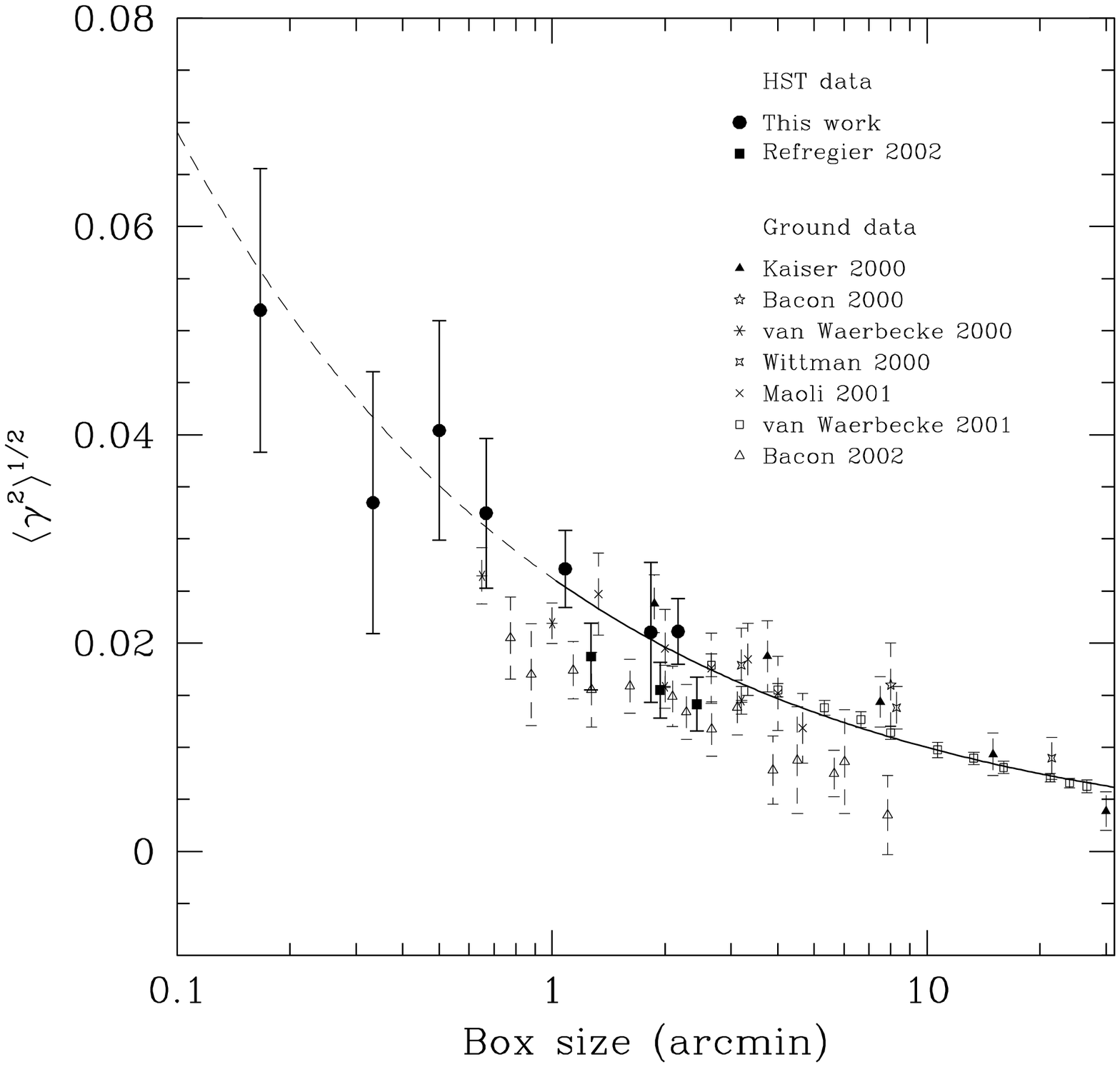}}
 \end{center}
 \figcaption{\figtwocap
 \label{fig:various shear measurements}}
 \end{inlinefigure}
 \fi

Measurements of the cosmic shear by various authors are shown in
Figure~\ref{fig:various shear measurements}; space-based results are
shown with solid error bars, ground-based are dashed.  The curve
represents the Jain \& Seljak (1997) predictions for a $\Lambda$CDM
cosmology with $ \sigma_8 = 0.6 $ and source redshift $ 1.0 $; the
functional approximation used is questionable below $ 1\arcmin $,
where the curve is dashed.  Several authors (e.g., Brown et al.\ 2003)
offer a more detailed comparison of the cosmic shear measurements with
model predictions, and discuss the more detailed constraints that can
be obtained from improved samples. Indeed, the statistics of
large-area ground-based measurements are beginning to approach the
regime in which specialized weight functions can be applied to measure
the strength of the density fluctuations on specific scales; see,
e.g., the discussion in Schneider et al.~(2002) and van Waerbecke et
al.\ (2002).

Among these results, space-based data maintain a unique value even when
compared with large-area, statistically impressive ground-based
measurements.  First, space-based measurements can probe smaller scales
than has been possible from the ground thus far; the small-angle regime
is uniquely sensitive to non-linear growth effects, and thus can serve
to discriminate between broadly similar cosmological models that differ
in the non-linear regime.  Second, space-based measurements are more
direct, requiring much less averaging and PSF correction; therefore they
suffer from different systematics than ground-based measurements, and
can be applied more easily to galaxies of small angular size.  Third,
the space can, at least at present, probe to fainter source magnitudes,
and thus in principle gain an understanding of the growth of structures
at higher redshift.  Note that, for example, Barber (2002) predicts a
significantly different scaling of the cosmic shear with both angular
scale and redshift than Jain \& Seljak (1997).  Our results are based on
a fairly large sample of HST data; however, the next few years will see
an explosive growth in the amount of data taken by HST that can be used
to measure cosmic shear, thanks to the Advanced Camera for Surveys and,
later, the Wide Field Camera 3.  We therefore expect a substantial
improvement in space-based measurements of cosmic shear, which will
probably help further constrain cosmology and the growth of dark-matter
structure in both the linear and non-linear regimes. 

\acknowledgments
 
This paper is based on observations with the NASA/ESA Hubble Space
Telescope, obtained at the Space Telescope Science Institute, which is
operated by the Association of Universities for Research in Astronomy,
Inc., under NASA contract NAS5-26555. The HST Archival research was
funded by STScI grant GO9212.

 \ifsubmode
  \clearpage
 \fi

\ifsubmode

\newpage

 \begin{inlinefigure}
 \begin{center}
 \resizebox{\textwidth}{!}{\includegraphics{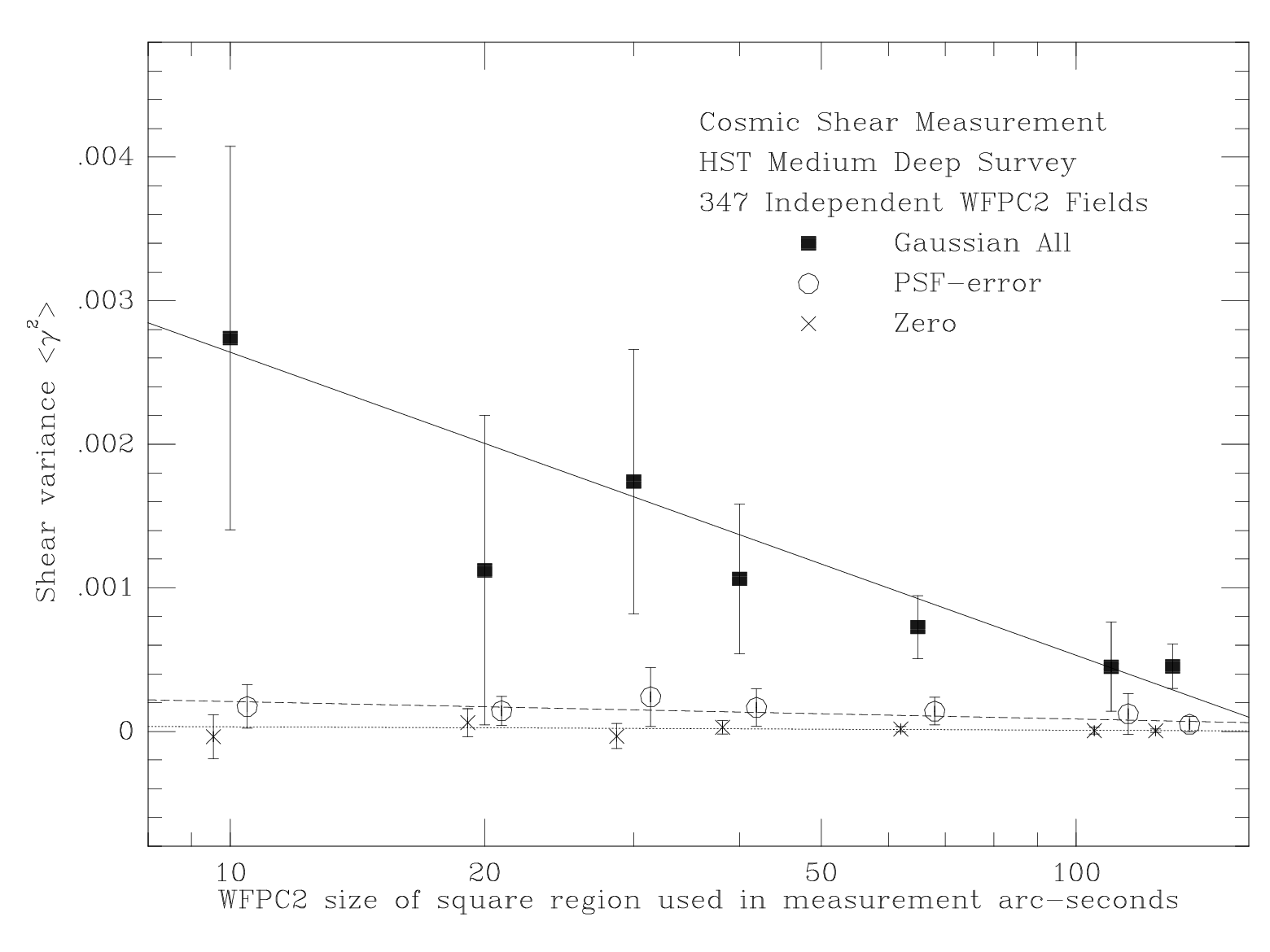}}
 \figcaption{\figonecap \label{fig:final shear}}
 \end{center}
 \end{inlinefigure}

\newpage

 \begin{inlinefigure}
 \begin{center}
 \resizebox{\textwidth}{!}{\includegraphics{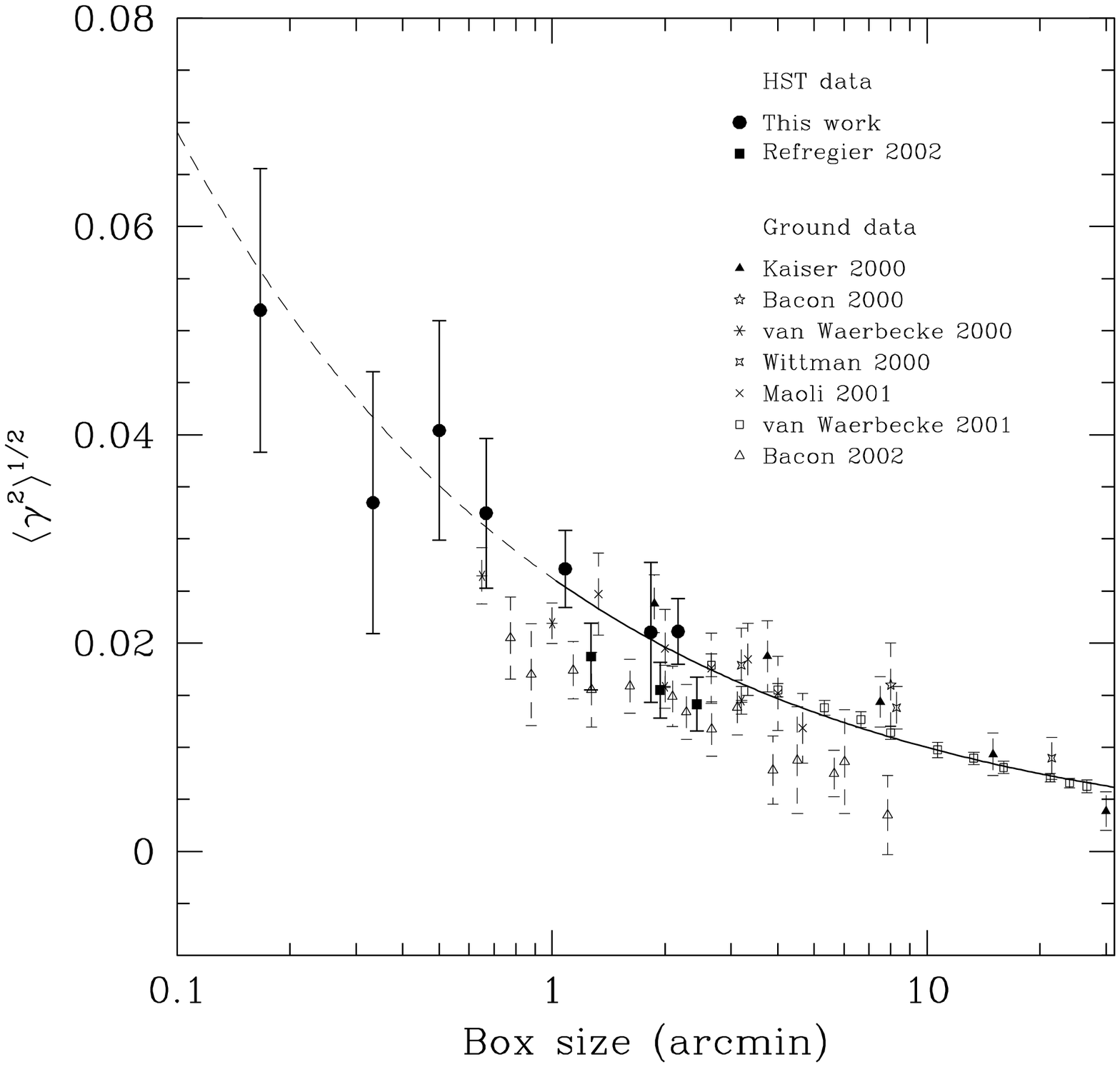}}
 \figcaption{\figtwocap \label{fig:various shear measurements}}
 \end{center}
 \end{inlinefigure}

\fi

 \end{document}